\documentstyle[aaspp4]{article}
\begin{document}

\title{$UBVI$ Color-Magnitude Diagrams in Baade's Window: Metallicity
Range, Implications for the Red Clump Method, Color ``Anomaly'' and
the Distances to the Galactic Center and the Large Magellanic
Cloud\footnote{Based on the observations collected at the Las Campanas
Observatory 2.5~m DuPont telescope}}

\author{K. Z. Stanek\altaffilmark{2}}
\affil{Harvard-Smithsonian Center for Astrophysics, 60 Garden St., MS20,
Cambridge, MA 02138}
\affil{\tt kstanek@cfa.harvard.edu}
\author{J. Kaluzny}
\affil{N.~Copernicus Astronomical Center, Bartycka 18, 
Warszawa PL-00-716, Poland}
\affil{\tt jka@camk.edu.pl}
\author{A. Wysocka}
\affil{Warsaw University Observatory, Al. Ujazdowskie 4,
PL-00-478 Warszawa, Poland}
\affil{\tt wysockaa@sirius.astrouw.edu.pl} 
\author{I. Thompson}
\affil{Carnegie Observatories, 813 Santa Barbara St., Pasadena, 
CA 91101-1292}
\affil{\tt ian@ociw.edu}
\altaffiltext{2}{Hubble Fellow}

\begin{abstract}

We analyze the $UBVI$ color-magnitude diagrams towards the Galactic
bulge in a relatively low-reddening region of Baade's Window. The
dereddened red giant branch is very wide [$\sim 1.0\;$mag in $(U-B)_0$
and $\sim 0.4\;$mag in $(B-V)_0$ and $(V-I)_0$], indicating a
significant dispersion of stellar metallicities, which by comparison
with the theoretical isochrones and data for Galactic clusters we
estimate to lie between approximately $-0.7< [Fe/H] < +0.3$,
i.e.~spanning about $1\;dex$ in metallicity, in good agreement with
earlier spectroscopic studies.

We also discuss the metallicity dependence of the red clump $I$-band
brightness and we show that it is between $0.1-0.2\;mag\;dex^{-1}$.
This agrees well with the previous empirical determinations and the
models of stellar evolution.

The de-reddened $(V-I)_0$ color of the red clump in the observed bulge
field is $\langle (V-I)_0 \rangle = 1.066 $, $ \sigma _{(V-I)_0} =
0.14$, i.e. $0.056\;$mag redder than the local stars with good
parallaxes measured by {\em Hipparcos}. It seems that the large
``color anomaly'' of $\sim 0.2\;$mag noticed by Paczy\'nski \& Stanek
and discussed in many recent papers was mostly due to earlier problems
with photometric calibration. When we use our data to re-derive the
red clump distance to the Galactic center, we obtain the
Galactocentric distance modulus $\mu_{0,GC}=14.69\pm 0.1\;$mag
($R_0=8.67\pm 0.4\;kpc$), with error dominated by the systematics of
photometric calibration.

We then discuss the systematics of the red clump method and how they
affect the red clump distance to the Large Magellanic Cloud. We argue
that the value of distance modulus $\mu_{0,LMC}=18.24\pm 0.08$
$(44.5\pm 1.7\;kpc)$, recently refined by Udalski, is currently the
most secure and robust of all LMC distance estimates. This has the
effect of increasing any LMC-tied Hubble constant by about 12\%,
including the recent determinations by the {\em HST}\/ Key Project and
Sandage et al.

The $UBVI$ photometry is available through the {\tt anonymous ftp}
service.

\end{abstract}

\keywords{Galaxy: center --- galaxies: distances and redshifts ---
galaxies: individual (LMC) --- stars: evolution --- stars:
horizontal-branch}

\section{INTRODUCTION}

The intermediate-age, degenerate core helium-burning stars, known as
the ``red clump'' stars, form a very pronounced and compact structure
on the color-magnitude diagrams (CMDs) in variety of stellar
populations.  Indeed the absolute magnitude-color diagram of the Solar
neighborhood obtained by {\em Hipparcos}\/ (Perryman et al.~1997)
clearly shows how compact the red clump is. Therefore, it could be
expected that this equivalent of the better known horizontal branch
stars in old, metal poor stellar populations, would be potentially a
very good standard candle. However, in spite of their large number and
good theoretical understanding these stars have seldom been used as
the distance indicators.  Stanek (1995) and Stanek et al.~(1994, 1997)
used these stars to map the Galactic bar.  Considering that the red
clump is the only distance indicator well calibrated with the {\em
Hipparcos}\/ trigonometric parallaxes (with $\sim 10^{3}$ red clump
stars with parallaxes better than 10\%), it becomes very important to
understand various possible systematics of this method.

As noticed by Paczy\'nski \& Stanek (1998), the peak $I$-band
brightness of the red clump is remarkably constant over a broad range
of $V-I$ colors.  This lack of correlation of red clump $I$-band
brightness with $V-I$ color was further confirmed by data from such
varied populations as the halo and globular clusters in M31 (Stanek \&
Garnavich 1998), field stars in the LMC (Udalski et al.~1998; Stanek,
Zaritsky \& Harris 1998) and the SMC (Udalski et al.~1998) and 
clusters in the LMC and the SMC (Udalski 1998b).  This was interpreted
as lack of strong dependence of the red clump $I$-band brightness on
the metallicity, a very desirable feature for any distance indicator.
This conclusion was however questioned, among others by Girardi et
al.~(1998). They made use of a large set of evolutionary tracks to
show that more massive clump models are systematically bluer than the
less massive ones. They argue that the color range spanned by the red
clump stars is largely caused by the dispersion of masses (and hence
ages), and not their metallicity. It is therefore important to be able
to distinguish between these two possibilities, which we attempt in
this paper using multiband $UBVI$ data for a stellar field in a low
reddening region of Baade's Window. As we show in this paper, the
large range in color of the bulge red clump stars is mostly due to
large metallicity range of these stars. We use this fact to put limits
on the metallicity dependence of the red clump $I$-band luminosity.

We describe the data and the deredenning procedure in Section 2.  In
Section 3 we compare the bulge data with the multiband observations of
Galactic clusters and with the theoretical isochrones, and we estimate
the metallicity range for the bulge. In Section 4 we discuss the
metallicity dependence of the red clump.  In Section 5 we discuss the
``color anomaly'' of the bulge red clump and derive the red clump
distance to the Galactic center. Finally, in Section 6 we discuss the
possible systematics of the red clump distance determination method
and argue it provides currently the most robust distance to the LMC.

\section{THE DATA AND THE DEREDDENING PROCEDURE}

\begin{figure}[t]
\plotfiddle{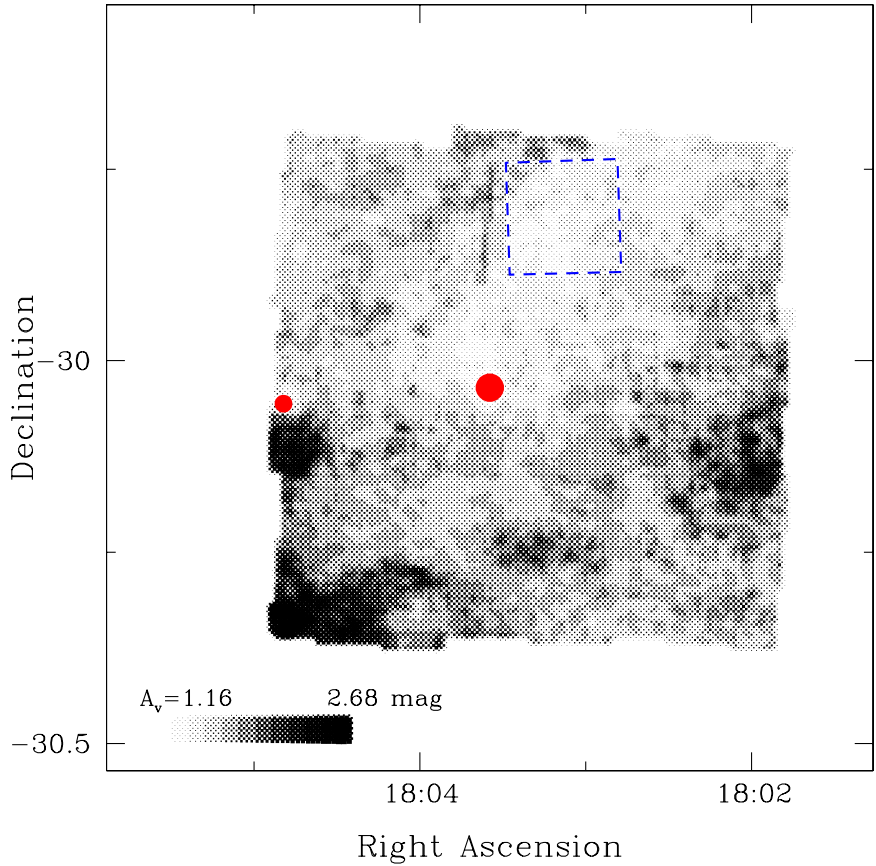}{10cm}{0}{60}{60}{-195}{-97}
\caption{$A_V$ extinction map of Baade's Window (Stanek 1996), along
with the BW8 field analyzed in this paper (dashed square).  The two
black dots correspond to the globular clusters NGC6522 (center) and
NGC6528 (left).}
\label{fig:map}
\end{figure}

The data were obtained with the Las Campanas Observatory 2.5-meter
DuPont telescope using a thinned Tektronix CCD -- known as TEK5 camera
-- with the pixel scale of $0.262\; arcsec\; pixel^{-1}$ and the field
of view $8.9\arcmin\times 8.9\arcmin$.  We observed a low-reddening
part of Baade's Window (Figure~\ref{fig:map}), falling within OGLE
field BW8 (Udalski et al.~1993).

Most of the data were obtained on the night of April 22/23, 1995 (UT).
A list of frames collected on that night and used to extract
photometry presented in this paper is given in Table~\ref{tab:log}.
Only $UBV$ filters were used during the April run.  However, the night
of April 22/23, 1995 turned out to be non-photometric, therefore on
the night of July 20/21, 1995 we collected single $UBVI$ frames (V-60
sec, B-60 sec, U-180 sec, I-40 sec) of the field offset $3.75\arcmin$
east relatively to the field observed in April. The same equipment was
used during both runs, with the exception of the B filter.  The
following relations were derived, by comparison with the Landolt
(1992) standard stars, to transform July photometry to the standard
$UBVI_{C}$ system:

\begin{eqnarray}
v = b_{1} + V -0.016\times (B-V) +0.144\times X\\
i = b_{2} + I -0.042\times (V-I) +0.05 \times X\\
b-v = b_{3} + 0.867\times (B-V) +0.128\times X\\
u-b = b_{4} + 0.956\times (U-B) +0.176\times X\\
v-i= b_{5} +1.057\times (V-I) +0.08\times X
\end{eqnarray}

\begin{figure}[p]
\plotfiddle{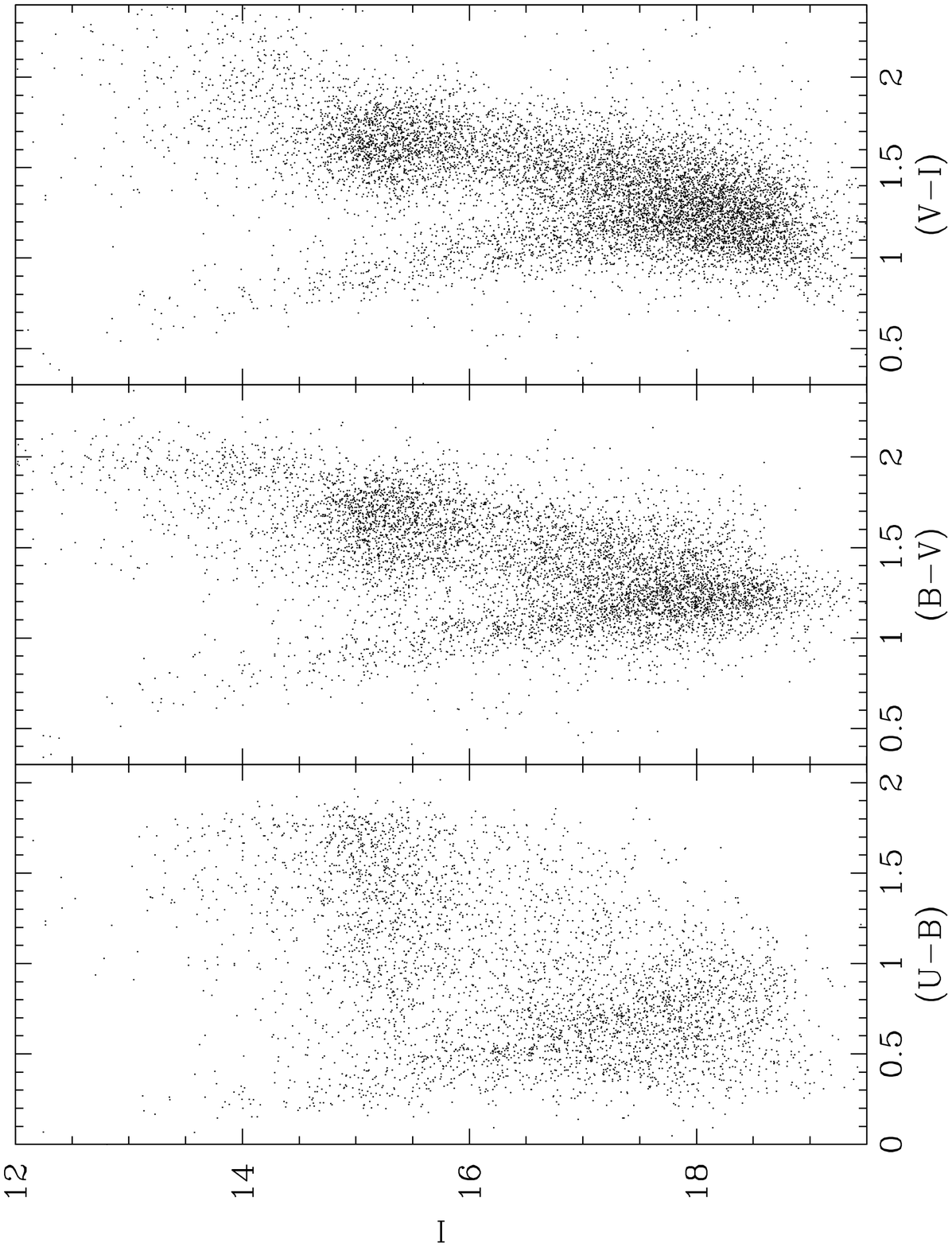}{10.5cm}{-90}{65}{65}{-270}{370}
\caption{$I$-band brightness vs. $(U-B), (B-V)$ and $(V-I)$.  Note
that the lack of red clump stars with $(U-B)>1.9$ is real and is not a
detection artifact.}
\label{fig:cmd}
\end{figure}

\tablenum{1}
\begin{planotable}{clc}
\tablewidth{15pc}
\tablecaption{\sc Observing log}
\tablehead{ \colhead{} & \colhead{$T_{exp}$} & 
\colhead{FWHM} \\  
\colhead{Filter} & \colhead{[$sec$]}  & 
\colhead{[$arcsec$]} }
\startdata  
$V$ & $3\times200$ & 1.00 \\
$V$ & $3\times200$ & 0.97 \\
$V$ & $1\times40$  & 0.79 \\
$B$ & $2\times500$ & 1.10 \\
$B$ & $3\times500$ & 0.86 \\
$B$ & $3\times500$ & 0.89 \\
$B$ & $1\times60$  & 0.89 \\
$U$ & $3\times900$ & 1.00
\enddata
\label{tab:log}
\end{planotable}

The $UBVI$ photometry obtained in July was used to establish zero
points of the $UBV$ data collected during the April run.  The
instrumental photometry was extracted using Daophot/Allstar package
(Stetson 1987). In case of the profile photometry we used gaussian
type empirical PSF varying quadratically with $X,Y$ coordinates.
Measurements with relatively large errors (for given magnitude) or
exceptionally large values of Daophot parameters CHI1,CHI2 were
flagged and not used for the analysis. The fraction of measurements
rejected for a given frame ranged from a few percent up to 65\%.  That
means that we sacrificed completeness of the photometry for the sake
of its internal accuracy.

We also derived equatorial coordinates for all stars, using the
USNO-A2.0 catalog (Monet et al.~1996), with an average difference of
$0.7\arcsec$ for the 215 transformation stars found in both
catalogs. The field center is at $[(\alpha,\delta)=
(18\!\!:\!\!03\!\!:\!\!08.1,-29\!\!:\!\!48\!\!:\!\!44), {\rm
J2000.0}]$. The $UBVI$ data and the coordinates of the stars can be
accessed from {\tt anonymous ftp} at {\tt
ftp://cfa-ftp.harvard.edu/pub/kstanek/BW8/}.

In Figure~\ref{fig:cmd} we plot the $I$-band brightness, for the stars
in the field BW8, versus all three independent colors $(U-B), (B-V)$
and $(V-I)$. The non-standard choice of plotting the $I$-band
brightness versus color is to show that the red clump $I$-band
brightness is basically independent of color. For more detailed
discussion of main features in the CMDs toward the Galactic bulge see
Kiraga, Paczy\'nski \& Stanek (1997).  The most striking feature of
the current CMDs (Figure~\ref{fig:cmd}) is the increasing width, in
color, of both the red clump and also the red giant branch, as we move
from the $(V-I)$ color to the $(U-B)$ color. The red clump and the red
giants branch become so wide in the $(U-B)$ color that they overlap
with the foreground Galactic disk stars.

We also use the data for two Galactic clusters for comparison
purposes.  The $BVI$ CMDs of the Galactic globular cluster 47Tuc were
taken from Kaluzny et al.~(1998)\footnote{Available at {\tt
ftp://www.astro.princeton.edu/kaluzny/Globular/47Tuc\_BVI/}}.  The
$UBVI$ data for an old open cluster NGC6791 were taken from Kaluzny \&
Udalski (1992) and from Kaluzny \& Rucinski (1995).

\begin{figure}[t]
\plotfiddle{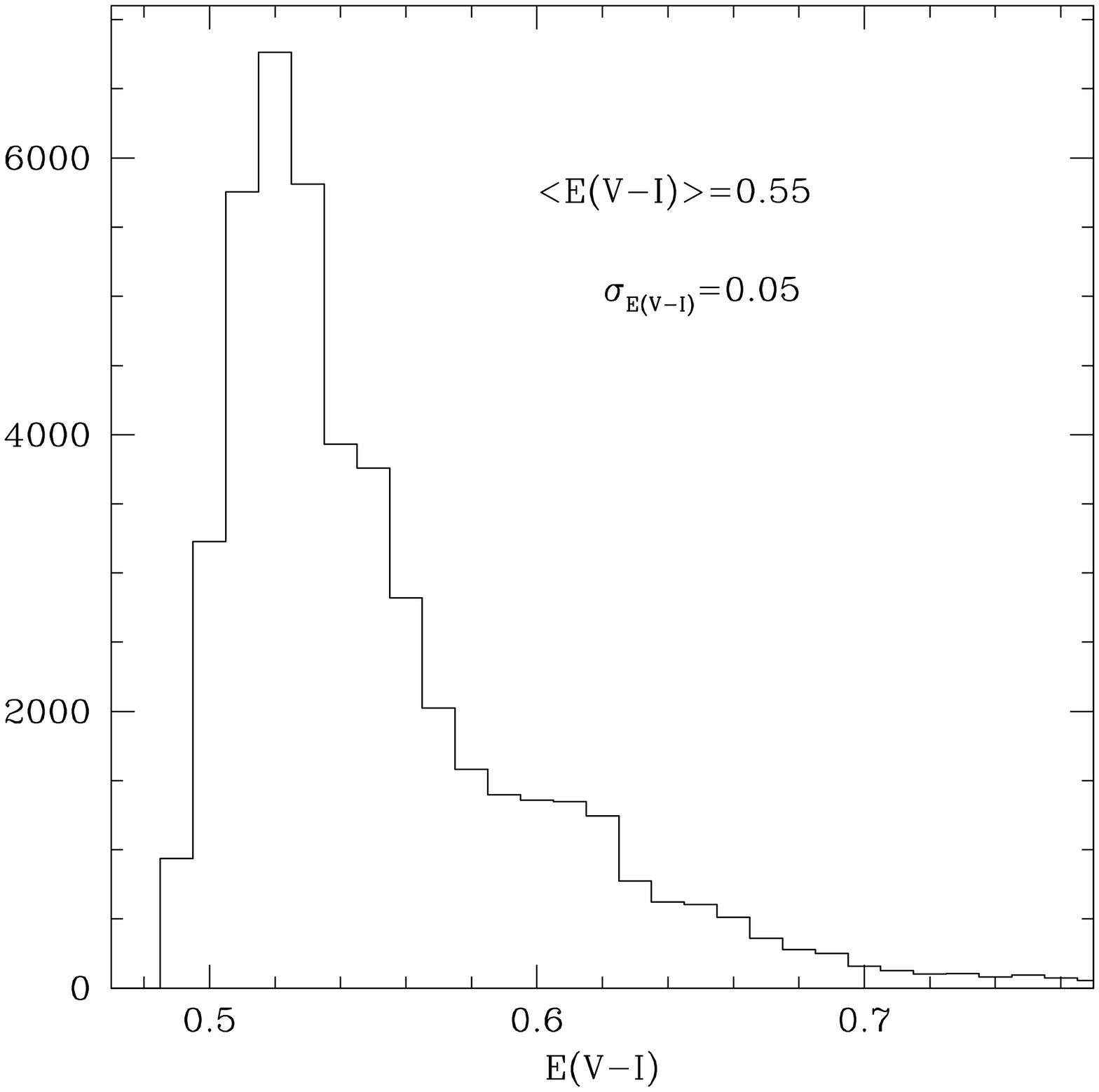}{8cm}{0}{50}{50}{-160}{-88}
\caption{Distribution of the $E(V-I)$ reddening for
stars located in the BW8 field, derived using the map of Stanek
(1996).  The reddening is fairly uniform across the field, with $rms$
scatter of $\sigma_{E(V-I)}=0.05\;$mag.}
\label{fig:evi}
\end{figure}

By design, we observed a relatively uniform- and low-reddening part of
Baade's Window, in order to minimize the effects of reddening in our
CMD studies.  To deredden the CMDs in the BW8 field, we used the
$A_V,E(V-I)$ extinction map of Baade's Window by Stanek
(1996),\footnote{Available at {\tt
ftp://www.astro.princeton.edu/stanek/Extinction/}} based on the method
of Wo\'zniak \& Stanek (1996) and the OGLE data of Udalski et
al.~(1993).  The zero point for Stanek (1996) map was determined by
Gould, Popowski \& Terndrup (1998) and Alcock et al.~(1998).  Figure
\ref{fig:evi} shows the distribution of the $E(V-I)$ reddening for
stars located in the BW8 field, derived using map of Stanek (1996).
The $E(V-I)$ reddening is quite uniform across the field, with small
$rms$ scatter of $\sigma_{E(V-I)}=0.05\;$mag. To remove the effects of
extinction in the $UB$ bands we used the standard coefficients, as
given for example by Schlegel, Finkbeiner \& Davis (1998, hereafter:
SFD), $A_B/A_V=1.324$ and $A_U/A_V=1.521$.

For the two clusters we have taken the reddening values from the SFD
map, $E(B-V)_{47Tuc}=0.031$ and $E(B-V)_{NGC6791}=0.15$.  In both
cases the SFD reddening values are very uniform over the clusters. In
case of 47Tuc the cluster is located at high galactic latitude of
$b=-44\arcdeg\!.9$, so the SFD value of reddening very likely
represents the total, low, reddening towards the cluster. In case of
NGC6791, located at low galactic latitude of $b=10\arcdeg\!.9$, it is
possible that some of the reddening measured by the SFD map could be
located behind the cluster. Indeed, in a recent detailed study of this
cluster by Chaboyer, Green \& Liebert (1999) they obtain the best-fit
value of $E(B-V)=0.10$, with the acceptable range of $0.08\leq E(B-V)
\leq 0.13$. On the other hand, NGC6791 is located at approximately
$4.2\;kpc$ from the Sun, which gives a distance from the Galactic
plane of about $800\;pc$, i.e. most probably well outside the Galactic
plane dust layer. We therefore assume for this cluster the SFD value
of $E(B-V)_{NGC6791}=0.15$, which has the advantage of being
independent from the color-magnitude data itself, and in fact is close
to some of the previous determinations [e.g. $E(B-V)=0.17$, Kaluzny \&
Rucinski 1995]. In any case, small changes in this value only weakly
affect any results of comparison with the Galactic bulge data
discussed in the next Section.

\begin{figure}[t]
\plotfiddle{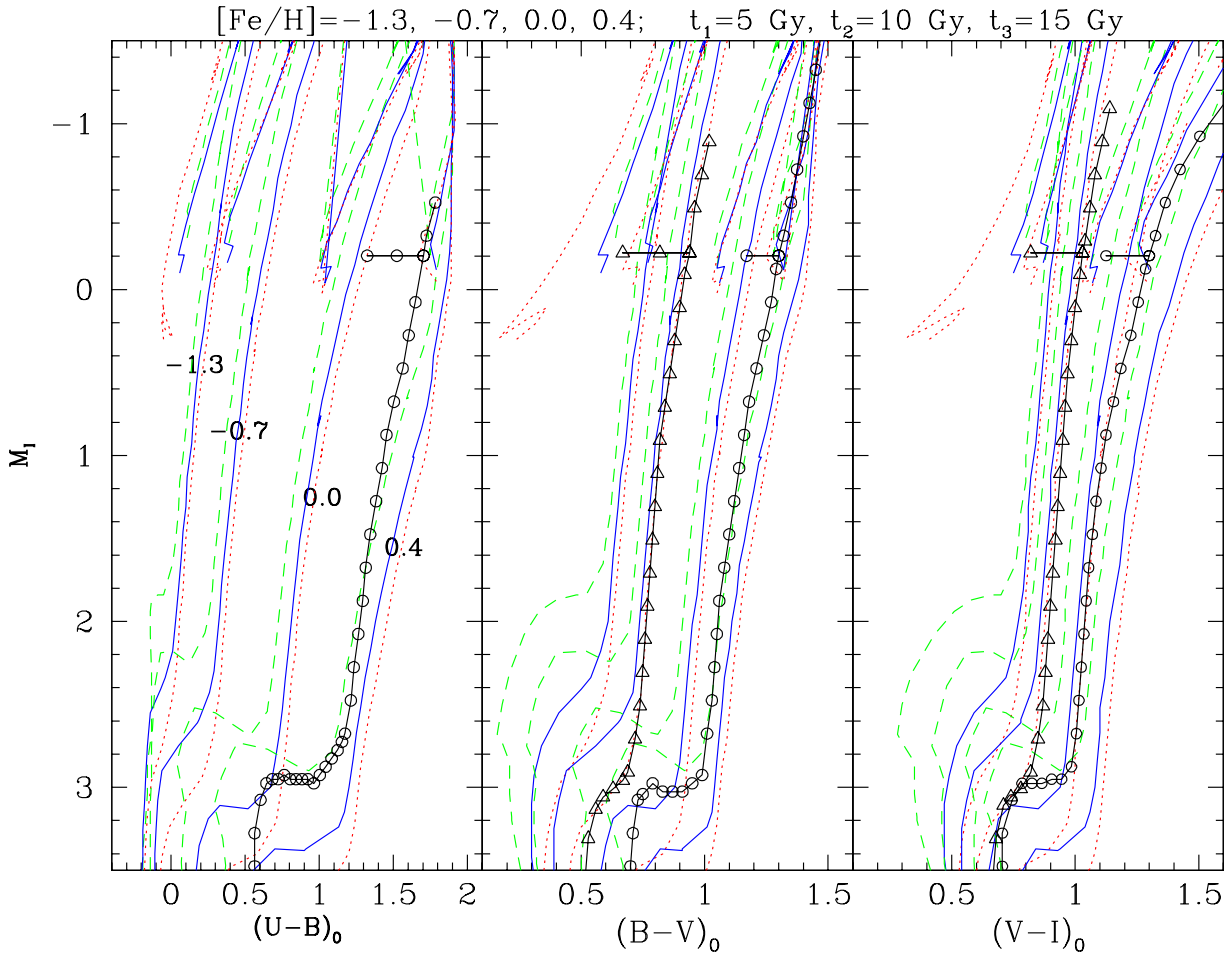}{11cm}{-90}{65}{65}{-270}{365}
\caption{Theoretical isochrones from Bertelli et al.~(1994), for four
metallicities: $[Fe/H]= -1.3, -0.7, 0.0$ and $+0.4$. For each
metallicity the isochrones corresponding to ages of 5.0, 10.0 and
$15.0\;Gyr$ are plotted ($5.0\;Gyr$: dashed lines; $10.0\;Gyr$:
continuous; $15.0\;Gyr$: dotted). Also shown are the dereddened CMDs
for 47Tuc ($[Fe/H]\approx -0.7$: triangles) and NGC6791
($[Fe/H]\approx +0.3$: circles). The red clump is represented by the
kink at $M_I\approx-0.2$ in the model isochrones.}
\label{fig:iso_i}
\end{figure}

\section{METALLICITY RANGE IN THE GALACTIC BULGE}

There have been several papers investigating spectroscopically the
abundance properties of the Galactic bulge, such as Rich (1988),
McWilliam \& Rich (1994), Minniti et al.~(1995) and Sadler, Rich \&
Terndrup (1996). They studied chemistry of giants in the Galactic
bulge, Rich (1998), McWilliam \& Rich (1994) and Sadler et al.~(1996)
in Baade's Window and Minniti et al.~(1995) in several other bulge
fields. From all these papers it is clear that while the mean
metallicity of the bulge is close to the solar value, at any given
field there is a large spread in metal abundances, possibly as large
as $2\;dex$ (see Figures 17 and 19 of McWilliam \& Rich, Figures 3-5
of Minniti et al. and Figure 11 of Sadler et al.). In this paper we
want to put additional constraints on the metallicity distributions,
using the multiband $UBVI$ photometric data. The spectroscopy is much
superior to photometry when it comes to metallicity investigations for
limited number of stars, but the photometry has the advantage of
providing some metallicity information for much larger samples of
stars.

We start by investigating the behavior of theoretical models of
stellar evolution, calculated by Bertelli et al.~(1994).  We select
four metallicities, corresponding to $[Fe/H]=-1.3, -0.7, 0.0$ and
$+0.4$. To illustrate the effect of age on the predicted colors and
luminosities, for each metallicity we take three isochrones,
corresponding to ages of 5.0, 10.0 and $15.0\;Gyr$, as to bracket the
most likely age for the population of the Galactic bulge
(e.g. Sevenster 1999).  In Figure~\ref{fig:iso_i} we plot the model
isochrones, showing $M_I$ luminosity as a function of $(U-B)_0,
(B-V)_0, (V-I)_0$ for each metallicity and age.

Also in Figure~\ref{fig:iso_i} we plot the dereddened CMDs (plotted as
maximum density ridge-lines only) for 47Tuc and NGC6791. In the recent
compilation Harris (1996) lists for 47Tuc a metallicity of about
$[Fe/H]\approx -0.7$.  Kaluzny et al.~(1998) has determined its
distance modulus to be $(m-M)_{0,47Tuc}=13.32\;$mag. For NGC6791
Chaboyer et al.~(1999) have derived a metallicity of $[Fe/H]=+0.4$,
somewhat higher than Kaluzny \& Rucinski (1995: $[Fe/H]=+0.3$) and
Garnavich et al.~(1994: $[Fe/H]=+0.2$), and distance modulus of
$(m-M)_{0,NGC6791}= 13.11\;$mag.  We adopt $[Fe/H]=+0.3$ for this
cluster. As we are going to use both these clusters only through their
RGBs, as a fiducial points for the metallicity determination, the
exact values of their distance moduli are not very important, since
the RGBs of the clusters are fairly vertical.  The adopted reddenings
are more important, but the differences are not significant from the
point of view of this paper.

As discussed by Girardi et al.~(1998), when modelling the {\em
Hipparcos}\/ CMD for the Solar neighborhood, the colors of the Bertelli
et al.~(1994) isochrones are systematically shifted to the redder
color. Indeed, to make the Bertelli et al.~(1994) isochrones agree
with the colors of the two clusters, we had to subtract $0.10\;$mag
from the theoretical $(B-V)$ color and $0.08\;$mag from the $(V-I)$
color.

\begin{figure}[t]
\plotfiddle{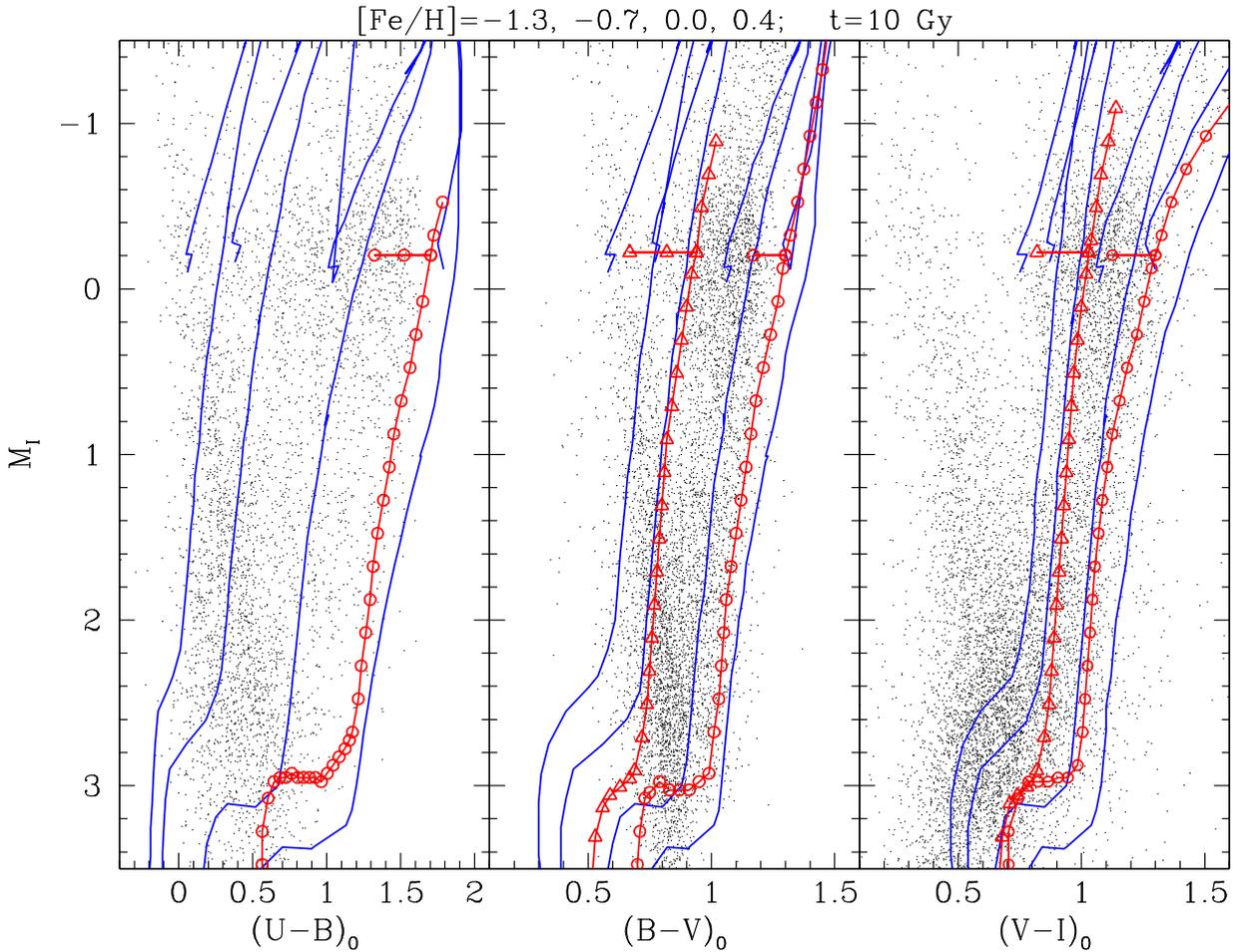}{11cm}{-90}{65}{65}{-270}{365}
\caption{$M_I$ absolute magnitude - dereddened color diagrams for the
BW8 field. Also plotted are the $t=10\;Gy$ isochrones and the the
dereddened maximum density ridge-lines for 47Tuc ($[Fe/H]\approx
-0.7$: triangles) and NGC6791 ($[Fe/H]\approx +0.3$: circles).}
\label{fig:cmd2}
\end{figure}

There is a number of interesting features in Figure~\ref{fig:iso_i}.
For example, the oldest and lowest metallicity isochrone
($[Fe/H]=-1.3, t=15\;Gy$) does not have a red clump, but has a clear
horizontal branch. However, all the remaining isochrones have well
defined red clump, with its $(U-B)_0$ and $(B-V)_0$ colors quite
strongly dependent on the metallicity, but only weakly dependent on
age.  However, it is worth noticing\footnote{We thank A. Udalski for
pointing that out to us.} that the $(V-I)_0$ color of the red clump is
basically the same for low metallicities, $(V-I)_0\approx 0.8$, both
from observations (Udalski et al.~1998) and from the theoretical
isochrones. The $I$-band brightness of the red clump, as given by the
isochrones, varies only little with age and metallicity. For each age,
higher metallicity corresponds to redder RGB (clearly seen between
$2.0<M_I<0.0$), while for each metallicity older age also corresponds
to redder RGB.  The model dependence of the RGB color on age is also
much weaker than the metallicity dependence, especially in the
$(U-B)_0$ color.

As the next step, using the RGB we want to constrain the metallicity
distribution for stars in our BW8 Galactic bulge sample. As discussed
above, the age effect on color is small, so for each metallicity we
will use the $t=10\;Gy$ isochrone as the representative one.  In
Figure~\ref{fig:cmd2} we plot the BW8 CMDs along with the theoretical
isochrones and the clusters data. The BW8 data has been dereddened
using the prescription described in the previous section, and also
shifted by distance modulus of $\mu_{0,GC}=14.57\;$mag determined by
Stanek \& Garnavich (1998).

As mentioned before, the RGB stars from the Galactic bulge overlap in
the $(U-B)$ color with the foreground Galactic disk main sequence
stars, and the full width of the red bulge giant branch is about $\sim
1.0\;$mag in this color. However, the RGB can be seen as a separate
feature when measured in $(B-V)$ and $(V-I)$ colors, with a full width
of about $\sim 0.4\;$mag. We will not attempt to derive a detailed
distribution in metallicity of the bulge RGB stars, as the current
sample of stars is not very large, and the colors of theoretical
isochrones are subject to problems discussed earlier in this
Section. Nevertheless, it is readily apparent that the metallicity
range of the bulge RGB stars extends approximately from that of 47Tuc
($[Fe/H]=-0.7$) to that of NGC6791 ($[Fe/H]=+0.3$), i.e. at least
$1\;dex$ in metallicity, with average metallicity close to the Solar
value. This agrees well with the spectroscopic determinations
discussed above (Rich 1988; McWilliam \& Rich 1994; Minniti et
al.~1995; Sadler, Rich \& Terndrup 1996).

It should be stressed here that the comparison of data with the
theoretical isochrones and CMDs of clusters serves only to establish a
rough metallicity range of the bulge stars. As discussed in
Paczy\'nski et al.~(1999), an observational program to establish how
good the correlation is between the metallicity and the colors would
be most useful. Such correlations for the local stars can be seen in
the first {\tt astro-ph} version of Udalski (2000).

\section{METALLICITY DEPENDENCE OF THE RED CLUMP}

As discussed by Paczy\'nski (1998) and Udalski (2000), the subject of
the metallicity dependence of the red clump brightness has important
consequences for the applicability of the method for the distance
measurements. Paczy\'nski \& Stanek (1998) have found that the
brightness of the red clump in the $I$-band does not depend on the
$(V-I)_0$ color over a broad range, which they expected to correlate
well with the metallicity.  However, Paczy\'nski (1998) has found no
such correlation between the $[Fe/H]$, obtained using Washington CCD
photometry (Geisler \& Friel 1992), and the color of the red clump
giants in the Galactic bulge.

Another approach to the question of the red clump metallicity
dependence comes from the population synthesis models.  Cole (1998),
in a very preliminary attempt, has derived a value for the metallicity
dependence of $0.21\pm 0.07\;mag\;dex^{-1}$, using models of Seidel,
Demarque \& Weinberg (1987), which by 1998 were already considered
obsolete (see discussion in Gibson 2000).  More sophisticated and
modern calculations of Girardi et al.~(1998) and Girardi (1999) give a
somewhat smaller value of $\sim 0.15\;mag\; dex^{-1}$ (this value, not
given explicitly by the author, was obtained from plots in Figures~4
and 7 of Girardi~1999). This should be compared to two empirical
determinations: $0.09\pm 0.03\;mag\; dex^{-1}$ by Udalski (1998a) and
more recent one of $0.13\pm 0.07\;mag\; dex^{-1}$ by Udalski (2000).
Popowski (2000) has re-analyzed the data of Udalski (1998a) and has
derived the red clump metallicity dependence of $0.19\pm 0.05\;mag\;
dex^{-1}$.  And while more work needs to be done on this problem, it
is already clear that the metallicity dependence of the red clump
brightness is modest ($0.1-0.2\;mag\; dex^{-1}$) and that theory and
observations are in a good agreement on the subject.

It is well worth noticing here that the data of Udalski (1998b) for
star clusters in the LMC and the SMC, used originally to show a weak
dependence of the red clump brightness on age, also provide reasonable
constrains on the red clump metallicity dependence. For example, the
SMC cluster L11 with $[Fe/H]=-0.70$ and age of $3.5\;Gyr$, has the
dereddened peak magnitude of the red clump only $0.03\;mag$ different
from another SMC cluster, NGC339, which has $[Fe/H]=-1.4$ and age of
$4.0\;Gyr$ (Da Costa \& Hatzidimitriou 1998). Even with relatively low
metallicity, NGC339 has a well-developed red clump, which agrees with
theoretical expectations (Figure~\ref{fig:iso_i}).

Additional constraint for the red clump metallicity dependence comes
from the data analyzed in this paper.  As we have shown in the
previous section, the bulge RGB spans about $1\;dex$ in metallicity,
from which follows naturally that the bulge red clump stars also have
a large metallicity range. At the same time, as noticed by Paczy\'nski
\& Stanek (1998), the peak $I$-band brightness of the red clump is
remarkably constant over a broad range of $(V-I)_0$ colors. In the
previous section we have shown that the model isochrones suggest a
good correlation between the metallicity and the color, and the data
presented in this paper support that strongly.  Paczy\'nski \& Stanek
(1998) have found a $0.09\;$mag difference in brightness between the
red clump stars with $1.0<(V-I)_0<1.1$ and those with
$1.3<(V-I)_0<1.4$ (their Fig.~1), with the redder stars fainter than
the bluer stars. Their red clump stars with $0.8<(V-I)_0<1.0$ were
however again fainter than those with $1.0<(V-I)_0<1.1$, but the bulge
red clump was not very well defined in this color range. If we take
this to represent the difference in brightness of red clump stars with
different metallicity, this gives us a small metallicity dependence of
$\sim 0.1\;mag\;dex^{-1}$, using the Galactic bulge metallicity range
derived in the previous section.  This is in good agreement with the
empirical determination of $0.13\pm 0.07\;mag\; dex^{-1}$ obtained by
Udalski (2000), and also in good agreement with the theoretical
expectations.

Paczy\'nski et al.~(1999) have presented $UBVI$ photometry in a nearby
field BWC in Baade's Window. Their results are overall in a
substantial agreement with ours, although there are some systematic
differences in the photometric zero points on the order of a few
hundredths of magnitude. These have no overall significance for the
main conclusions presented in this paper and in the paper of
Paczy\'nski et al.~(1999). One aspect which is affected, the ``color
anomaly'' of the red clump, is discussed in the next section.

\section{``COLOR ANOMALY'' OF THE RED CLUMP AND THE DISTANCE TO THE
GALACTIC CENTER}

As first noticed by Paczy\'nski \& Stanek (1998), the dereddened red
clump observed by OGLE-I in Baade's Window was $\sim 0.2\;$mag redder
than in the Solar neighborhood. Were this ``color anomaly'' real, it
would have important consequences, as extensively discussed by
Paczy\'nski (1998), Paczy\'nski et al.~(1999), Popowski (2000),  Alves
(2000) and Gould, Stutz \& Frogel (2000).

\begin{figure}[t]
\plotfiddle{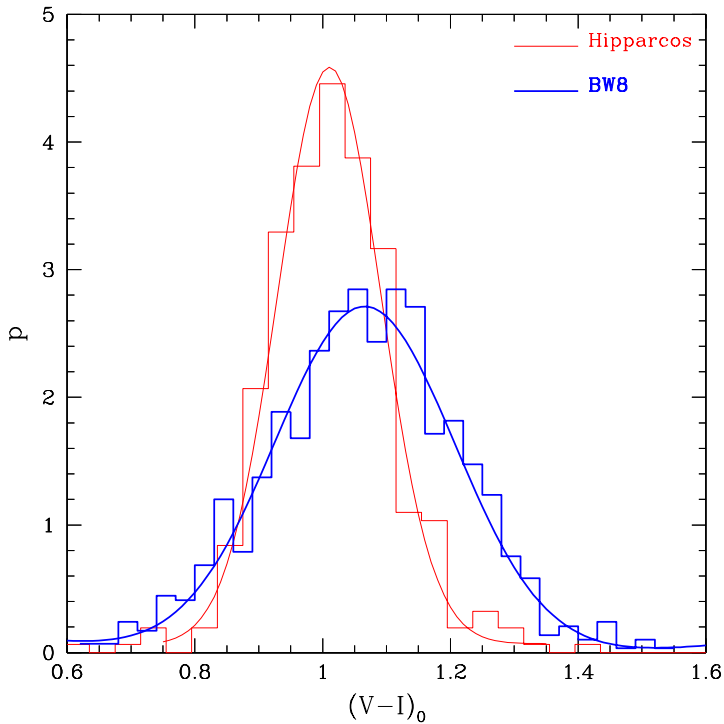}{8cm}{0}{50}{50}{-160}{-88}
\caption{The number of red clump stars in the solar neighborhood,
based on {\em Hipparcos}\/ data, is shown as a function of their $(V-I)$
color with the thin solid line. The number of red clump stars in the
BW8 field is shown as a function of their $(V-I)_0$ color with the
thick solid line.  All distributions are normalized, and include all
stars with their absolute magnitude within $0.3\;$mag of $M_{I,m}$.}
\label{fig:color}
\end{figure}

We have therefore decided to repeat the procedure of Paczy\'nski \&
Stanek (1998) and derive the dereddened $(V-I)_0$ color of the red
clump in the bulge. The results can be seen in Figure~\ref{fig:color},
analogous to Figure~4 of Paczy\'nski \& Stanek (1998).  The
de-reddened $(V-I)_0$ color of the red clump in the Galactic bulge is
$\langle (V-I)_0 \rangle = 1.066 $, $ \sigma _{(V-I)_0} = 0.14$,
i.e. $0.056\;$mag redder than the local stars with good parallaxes
measured by {\em Hipparcos}.  However, huge ``color anomaly'' of $\sim
0.2\;$mag seen by Paczy\'nski \& Stanek (1998) is mostly eliminated.
The remaining difference of $0.056\;$mag could be real: the chemical
composition of the bulge stars is somewhat different from the local
stars, although there is substantial overlap (see Figure~1 of Udalski
2000). Given that the main difference between the analysis of
Paczy\'nski \& Stanek (1998) and the current paper is in the different
photometric data used, we suspect that most of the ``color anomaly''
resulted from CCD non-linearity of OGLE-I data (Kiraga et al.~1997)
and resulting problems with photometric calibration. The fact that the
new OGLE-II data give $(V-I)_0=1.11$ (Paczy\'nski et al.~1999),
i.e. $\sim 0.1\;$mag smaller ``color anomaly'', support this
conclusion (see discussion in Popowski 2000).

Given a likely problem with the OGLE-I photometric data, which were
used by Paczy\'nski \& Stanek (1998), we decided to re-derive the red
clump distance to the Galactic center with current data.  We have
followed exactly the procedure of Paczy\'nski \& Stanek (1998): we
selected the red clump stars in the color range $0.8<(V-I)_0<1.25$ in
the BW8 field and we fitted the resulting distribution with their
Eq.1. The fitted Gaussian peak of the red clump distribution was at
$I_{0,m}=14.447\pm0.02$ (statistical error only), $0.12\;$mag fainter
than that of Paczy\'nski \& Stanek. This, combined with corrections
due to the geometry of the Galactic bar (Stanek et al.~1997) and
revised $I$-band calibration of the local red clump (Stanek \&
Garnavich 1998), gives the Galactocentric distance modulus
$\mu_{0,GC}= 14.69\pm 0.1\;$mag, with corresponding Galactocentric
distance $R_0=8.67\pm 0.4\;kpc$. We set the error to fairly
conservative $0.1\;$mag to allow for possible errors in photometric
calibration of current data.

This value can be compared to two recent determinations. The first
one, by Alves (2000), uses the infrared $K$-band version of the red
clump method, and is therefore much less susceptible to errors in the
reddening correction. His value is $R_0=8.24\pm 0.42\;kpc$, with the
error dominated by small number of bulge red clump giants with good
$K$-band photometry. Another determination comes from Gould et
al.~(2000), who obtained $R_0=8.63\pm 0.16\;kpc$ (statistical error
only).  This was obtained by using the value for the $I$-band peak of
the red clump from Paczy\'nski et al.~(1999), but after it was
adjusted for, different from Stanek (1996), value of the ratio of
total to selective extinction $R_{VI}$ (see Gould et al.~2000 for
discussion). Note that the value of Gould et al. is extremely close to
that obtained in this Section, but even the lower value of Alves
(2000) is in good statistical agreement and within 5\% of our value.

\begin{figure}[p]
\plotfiddle{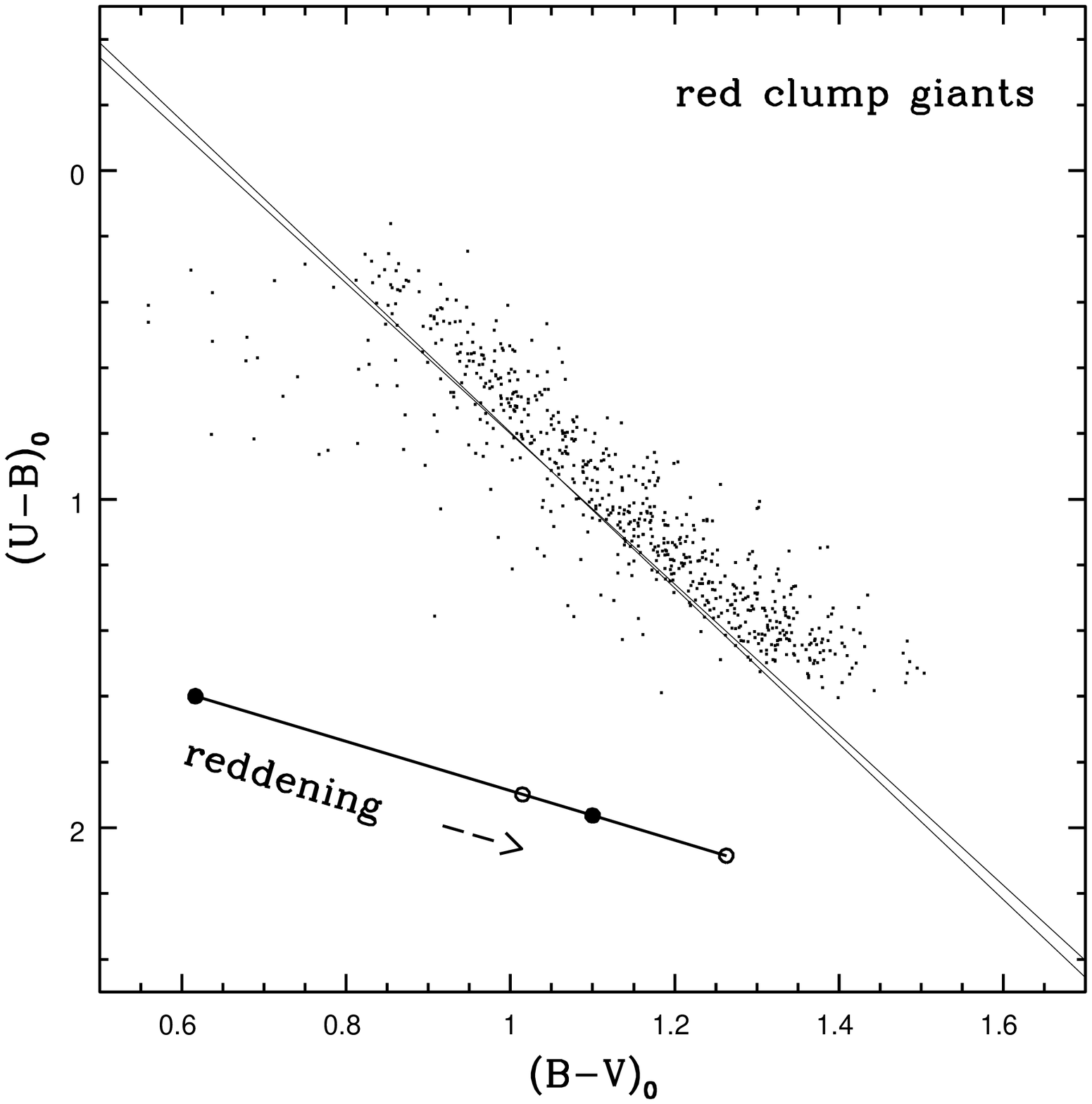}{7.5cm}{0}{50}{50}{-160}{-81}
\caption{The color -- color diagram: $(B-V)_0 - (U-B)_0$ for red clump
giants in the BW8 field, corrected for interstellar reddening.  The
two thin lines that cross are the regression lines of one color with
respect to another, derived by Paczy\'nski et al.~(1999) from the
local stars with {\em Hipparcos}\/ parallaxes better than 10\%.}
\label{fig:ubbv}

\plotfiddle{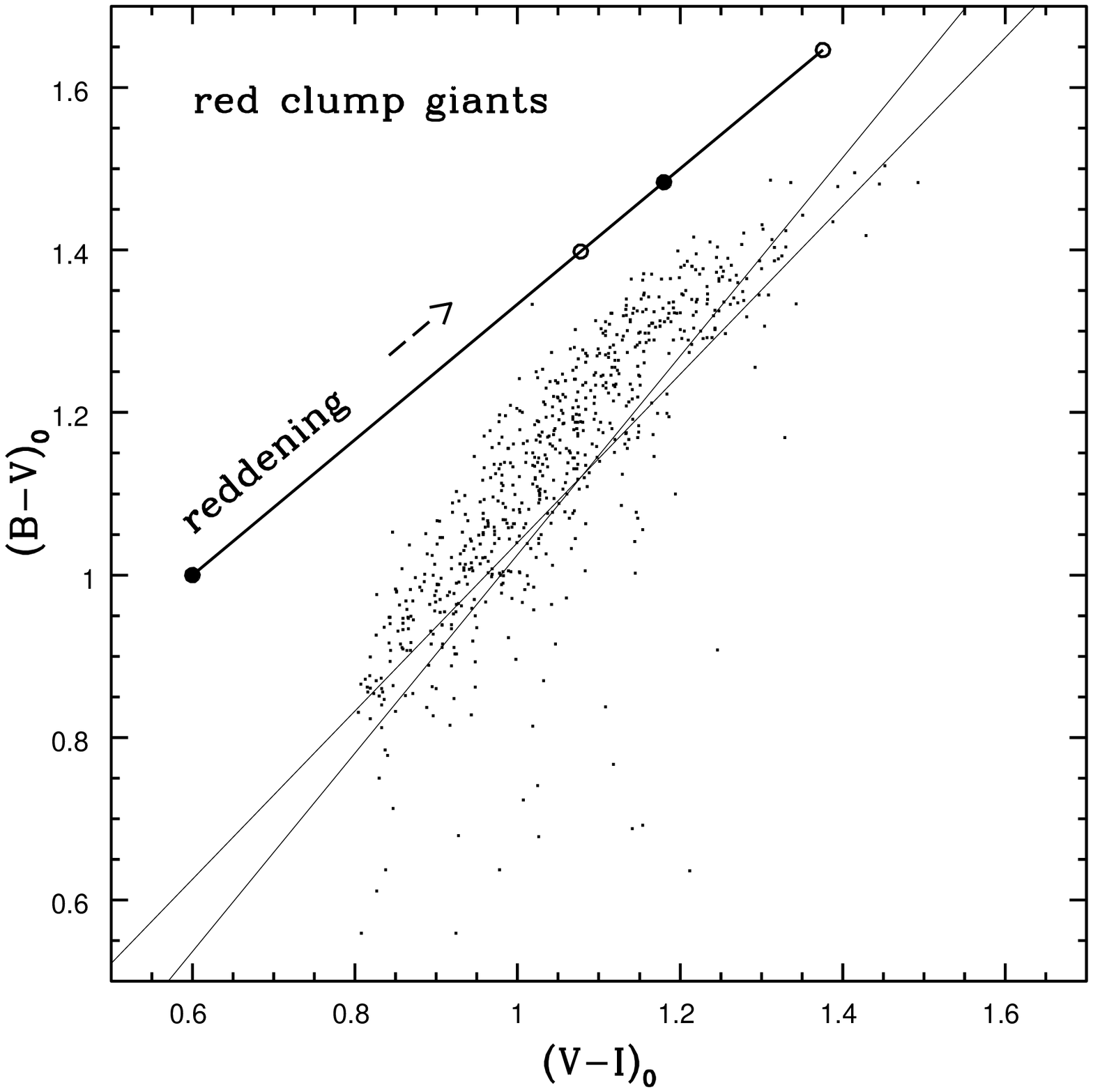}{8.7cm}{0}{50}{50}{-160}{-83}
\caption{The color -- color diagram: $(V-I)_0 - (B-V)_0$ for red clump
giants in the BW8 field, corrected for interstellar reddening.  The
two thin lines that cross are the regression lines of one color with
respect to another, derived by Paczy\'nski et al.~(1999) from the
local stars with {\em Hipparcos}\/ parallaxes better than 10\%.}
\label{fig:bvvi}
\end{figure}

To further explore the issues related to photometric calibration,
following Paczy\'nski et al.~(1999) we decided to investigate the
color-color properties of our sample of red clump giants.  This was
done exactly as in Paczy\'nski et al.~(1999) paper, and the results
are presented in Figures~\ref{fig:ubbv} and \ref{fig:bvvi}, analogous
to Figures~6 and 7 in their paper.  The sequences defined by the bulge
giants are fairly close to those defined by the local stars, although
there seems to be a small systematic offset, possibly due to small
errors in the photometric calibration.

It is clear from the above discussion that the issue of accurate
photometric calibration is of crucial importance in these extremely
crowded Galactic bulge fields. In addition, due to fairly large
reddening, the exact properties of interstellar extinction need to be
better understood (see Gould et al.~2000 for discussion of one of the
possible problems).

\section{RED CLUMP SYSTEMATICS AND THE DISTANCE TO THE LMC}

The distance to the LMC is one of the most important problems of the
modern astrophysics, because the extragalactic Cepheid-based distance
scale is tied to it (e.g. Mould et al.~2000). Most reviews of the
subject place the LMC at the distance modulus of $\sim 18.5\pm
0.1\;$mag, but values as high as $\sim 18.7$ are also present in the
literature (Feast 1999). The errors in this distance are quoted
sometimes as being as small as $0.05\;$mag (e.g. van den Bergh 2000,
Carretta et al. 2000), although these small error bars are obtained by
statistically unjustified procedures of ignoring strong outliers or
assigning them arbitrarily large error bars. It should be noted that
averaging various distance estimates while they are dominated by their
systematic errors is a risky statistical proposition.  For a good
discussion of the distance to the LMC derived from different methods
see Jha et al.~(1999) and Gibson (2000).

As discussed throughout this paper, the systematics of the red clump
method seem to be quite well understood, both empirically and
theoretically.  In the rest of this Section we will present a summary
of various factors affecting the robustness of the red clump distance
to the LMC and compare it with other commonly used methods.

\paragraph{Zero point calibration}  As for any other distance
indicator, it is important to know the brightness of the red clump for
some selected stellar population. This was provided by the {\em
Hipparcos}\/ satellite, which measured $\sim 10^3$ red clump stars with
better than 10\% parallaxes (Perryman et al.~1997).  Stanek \&
Garnavich (1998) selected from them a sub-sample of 228 red clump
stars with the distance $d<70\;pc$.  They found that the red clump
$I$-band absolute magnitude $M_{I,m}(d<70\;pc) = -0.227 \pm 0.023$ for
these nearby stars, with the average distance $\langle d_{<70}\rangle
=50\;pc$.  One expects very little reddening, $E(B-V)<0.02$, for such
nearby stars, so they assumed that their $d<70\;pc$ sample suffered no
reddening. As follows from detailed discussion of the local
interstellar medium by Ferlet (1999), this assumption is unlikely to
produce any significant ($>0.05\;mag$) error --- we live in the Local
Bubble, largely devoid of stellar extinction. This assumption can be
further tested by comparing the brightness of various spatial
sub-samples of the red clump, e.g. low-$b$ vs. high-$b$ samples.

It is worth mentioning that Girardi et al.~(1998) have applied the
Lutz-Kelker (1973) bias correction while calculating the value of
$M_{I,m}$ for the {\em Hipparcos}\/ red clump and have obtained a value
of $M_{I,m}=-0.21$, i.e. practically identical to that of Stanek \&
Garnavich (1998).

It should be stressed that {\em Hipparcos}\/ provided accurate
distance determinations for over 1,000 red clump stars, but
unfortunately $I$-band photometry is available for only $\sim 50\%$ of
them, so it would be important to obtain $I$-band photometry for all
{\em Hipparcos}\/ red clump giants. This would further reduce the
already very small statistical zero-point error and would allow better
testing of any systematic errors.

It is beyond the scope of this paper to provide detailed reviews of
the zero-point calibrations for the other commonly used distance
indicators, so the interested reader should consult Popowski \& Gould
(1999) for the RR Lyrae stars, Carretta et al.~(2000) for the subdwarf
fitting technique and Feast (1999) for the Cepheids zero point (or one
of several other reviews).

To provide some comparison with the red clump {\em Hipparcos}\/ zero
point calibration, based on many hundreds of stars measured with
better than 10\% parallaxes, there is only {\em one} RR Lyrae with
reasonable trigonometric parallax (RR Lyrae itself: Popowski \& Gould
1999), similarly for the Cepheid variables (Polaris: Feast \&
Catchpole 1997). There is somewhat more {\em Hipparcos}-measured
subdwarfs, from about 15 in Reid (1997) to about 30 in Carretta et
al.~(2000). It is fair to say that the red clump stars are the only
well represented distance indicator in the sample of Solar
neighborhood stars with trigonometric parallaxes well measured by {\em
Hipparcos}.

\paragraph{Metallicity dependence} Again, as with any distance indicator,
also red clump stars are subject to possible metallicity dependence.
We have reviewed this problem throughout this paper, so let us stress
again that this dependence for the red clump is modest ($\sim
0.1-0.2\;mag \; dex^{-1}$) and the empirical determinations agree well
with the model calculations. This is quite unlike the Cepheids, where
the empirical determinations range from 0 to $-0.4\;mag\; dex^{-1}$
(Freedman \& Madore 1990; Sasselov et al.~1997; Kochanek et al.~1997;
Kennicutt et al.~1998), while the theoretical determinations often
have even different sign of the metallicity effect (e.g. Gautschy
1998). The situation for the metallicity dependence of RR Lyrae stars
seems to be not so controversial, and the interested reader should
consult any of the many recent papers on the subject (e.g. Popowski \&
Gould 1999). However, the metallicity is of basic importance for the
subdwarf fitting method, where the metallicity of faint, locally
observed subdwarfs is compared to metallicity of giants in distant
globular clusters (for discussion see Popowski \& Gould 1999). So, it
is again fair to say that if the red clump method has any problems
with the metallicity dependence, for other local Group distance
indicators these problems are similar or in some cases much more
severe.

\paragraph{Age dependence} It is possible that red clump stars would 
suffer from a strong age dependence. This possibility was however not
confirmed by empirical determination of Udalski (1998b), who observed
star clusters with different ages in the LMC and the SMC and found
basically no age dependence between 2 and $10\;Gyr$. Also, modern
theoretical models of Girardi et al.~(1998) and Girardi (1999) do not
support strong age dependence.

Recently, Sarajedini (1999) found that the $M_I$ peak brightness of
the red clump in Galactic open clusters becomes fainter for clusters
older than $5\;Gyr$. However, there are only three such clusters in
his sample, so much more observational work needs to be done to firmly
establish age dependence, if any.  More importantly, the absolute peak
brightness $M_I$ for each cluster is derived by Sarajedini (1999)
using main sequence fitting technique, which is not without its own
problems (e.g. Pinsonneault, Terndrup \& Yuan 2000). Finally, even the
cluster with the brightest red clump in Sarajedini's sample, NGC 2204,
has $M_I=-0.34$ (at $[Fe/H]=-0.34$), i.e. only $0.11\;$mag brighter
than the local {\em Hipparcos}\/ red clump (Stanek \& Garnavich 1998),
which gives an approximate size of population correction to the red
clump LMC distance. Obviously, population effects do exist for the red
clump method, but they are fairly small.

Possibly the strongest argument against significant age dependence of
the red clump method, when applied to mixed populations of stars,
comes from Girardi (2000). He noticed that for constant star formation
rate the age distribution of clump stars is strongly biased towards
intermediate ages, $1-3\;Gyr$ (his Fig.~2). This very strongly suggest
that when deriving the red clump distance to objects like the LMC one
largely bypasses the age problem because the majority of the red clump
stars in both the local and the LMC populations are of similar age.
This conclusion is also supported by the very compact red clump in the
SMC, as presented by Paczy\'nski et al.~(1999, their Figure~10).

\paragraph{Reddening} Every method of distance determination 
utilizing standard candles is affected to some extent by how well the
reddening and extinction along the line of sight are known. In this
aspect red clump is not worse than any other method, and since it
relies on the $I$-band brightness, it might be somewhat less
susceptible to reddening than some other methods. In many cases where
the red clump was used to derive the distance to the LMC (Udalski et
al.~1998; Stanek et al.~1998; Udalski 1998a,b), particular attention
was paid to come up with as accurate estimate of the reddening as
possible. For example, in the paper of Stanek et al.~(1998) the
reddening values obtained using the reddening map of Harris, Zaritsky
\& Thompson (1997) were compared to those from the SFD map and in the
two regions selected for further analysis the two maps agreed to
within $0.02\;$mag in $E(B-V)$, or $0.04\;$mag in $A_I$.

Still better approach to the problem of reddening is to apply any
distance determination method in regions with small reddening, if
possible. Udalski (2000) used nine stellar fields in the LMC to derive
the distance modulus of $18.24\pm 0.08$. The SFD map was used to
obtain the reddenings, and some of the LMC fields used in his paper
have reddening values as small as $E(B-V)\approx 0.03$ (see Table 1 of
Udalski 2000).  The approach of bypassing the reddening problem by
using low extinction regions is, by design, much more preferred over
complicated dust corrections applied in dusty regions (Zaritsky 1999;
Romaniello et al.~2000; Sakai, Zaritsky \& Kennicutt 2000).  Another
approach, which would eliminate most of the problems due to the
reddening, would be to apply the $K$-band version of the red clump
method (Alves 2000) to obtain the LMC distance.

To summarize, the major strength of the red clump distance
determination technique lies in its accurate absolute magnitude as
determined in the Solar neighborhood using {\em Hipparcos}\/ data
(Paczy\'nski \& Stanek 1998; Stanek \& Garnavich 1998). It was also
shown the the $I$-band brightness of the red clump remains remarkably
constant in such varied environments as the halo and globular clusters
of M31 (Stanek \& Garnavich 1998), field stars in the LMC and the SMC
(Udalski et al.~1998; Stanek et al.~1998) and clusters in the LMC and
the SMC (Udalski 1998b). Also, as shown by Paczy\'nski \& Stanek
(1998), the $I$-band brightness of the Galactic bulge red clump varies
only by few hundreds of magnitude over $\sim 0.4\;$mag in
$(V-I)_0$. This and analysis of Udalski (1998a) and Udalski (2000)
indicate that there is only a modest metallicity dependence of the red
clump $I$-band brightness, in agreement with the theoretical models
(Girardi 1999).  Also, there is no significant age dependence over a
broad range of ages (Udalski 1998b). All this combined makes the
distance modulus to LMC of $18.24\pm 0.08$, as determined by Udalski
(2000), currently the most secure and robust of all LMC distance
estimates, which has the effect of increasing any LMC-tied Hubble
constant determinations by about 12\%, including the recent
determinations by the {\em HST}\/ Key Project (e.g. Mould et al.~2000)
and by Sandage \& Tammann (1998).

\paragraph{NOTE TO AVOID CONFUSION:} After the first version of this
paper was submitted for publication in AJ (August 1999:
astro-ph/9908041), there appeared several papers discussing the
properties of red clump. Some of these papers cite this paper, but are
also cited in this current revised version.

\acknowledgments{We thank Bohdan Paczy\'nski, Andrzej Udalski and
Peter Garnavich for many most useful and interesting discussions and
very helpful comments on the manuscript. We also thank the anonymous
referee for helpful comments on the paper, and Bohdan Paczy\'nski for
help in preparing two of the figures. KZS was supported by the
Harvard-Smithsonian Center for Astrophysics Fellowship and by NASA
through Hubble Fellowship grant HF-01124.01-99A from the Space
Telescope Science Institute, which is operated by the Association of
Universities for Research in Astronomy, Inc., under NASA contract
NAS5-26555. JK was supported by NSF grant AST-9819787 to Bohdan
Paczy\'nski and by the Polish KBN grant 2P03D003.17.}

\end{document}